\documentclass[proof]{WileyASNA-v2}

\articletype{Article Type}%

\received{1 October 2020}
\revised{}
\accepted{}

\begin{document}

\title{General relativistic formulas for mass and spin of a Kerr black hole in terms of redshifts and orbital parameters }

\author[1]{Alfredo Herrera--Aguilar}

\author[2]{Ra\'ul Lizardo--Castro}

\author[3]{Ulises Nucamendi}

\authormark{Alfredo Herrera--Aguilar \textsc{et al}}

\address[1]{\orgdiv{Instituto de F\'isica}, \orgname{Benem\'erita Universidad Aut\'onoma de Puebla (BUAP)}, \orgaddress{\state{Puebla}, \country{ M\'exico}}}

\address[2]{\orgdiv{Instituto de Ciencias Nucleares}, \orgname{Universidad Nacional Aut\'onoma  de M\'exico  (UNAM)}, \orgaddress{\state{Mexico City }, \country{ M\'exico}}}

\address[3]{\orgdiv{Instituto de F\'{\i}sica y Matem\'{a}ticas}, \orgname{Universidad Michoacana de San Nicol\'as de Hidalgo (UMSNH)}, \orgaddress{\state{Michoac\'an }, \country{ M\'exico}}}

\corres{{Av. San Claudio y 18 Sur, Col. San Manuel, Ciudad Universitaria, CP 72570 Puebla, Pue., M\'exico}  \email{aherrera@ifuap.buap.mx}}

\abstract{We derive closed formulas for the mass $M$ and spin $a$ parameters of a Kerr black hole in terms of a minimal quantity of observational data: the red-/blue-shifts of photons emitted by massive particles (stars) moving on geodesics around the black hole and their respective orbital radius. It turns out that given a set of two (three) stars revolving around the black hole, these formulas involve just eight (twelve) observational data. For the case of a single star orbiting the black hole we need a minimal set of four observational measurements to analytically determine both parameters. }


\maketitle

\section{Introduction}
\label{intro}

During last decades we have witnessed a growing interest in the search for astrophysical black holes, whose existence has become a fundamental scientific issue, as well as for different methods to characterize their parameters \citep{Shen,Ghez,Morris,Eckart,Dokuchaev}. On the one hand there is a vast dynamic evidence indicating that in each galaxy there are millions of black holes with stellar masses; besides, at the center of almost all galaxies there exists a supermassive black hole (with masses that range from millions to billions of solar masses), including a black hole hosted at the center of the Milky Way, called SgrA* \citep{Begelman}. On the other hand, recent discoveries of gravitational waves \citep{GWs1,GWs2,GWs3,GWs4} have also given rise to a new golden era in the area of black hole astrophysics since in the near future more sensitive detectors will be built, starting the physics of high precision gravitational waves that will allow probing regions increasingly closer to their event horizons \citep{gwbhreview}. Moreover, the recent report on the  {\it shadow} of the black hole hosted at the center of the giant elliptical galaxy M87 \citep{EHT1,EHT4,EHT6}, which points to another confirmation of the predictions of the general theory of relativity, has strengthen as well the interest in these mysterious objects.

Regarding the center of our Galaxy, the GRAVITY collaboration has recently reported the detection of the Schwarzschild precession in the orbit of the star S2 and has estimated the mass of SgrA* to be $M = 4.261\times 10^{6} M_{\bigodot}$ and the distance from the Earth to its center $R_0 = 8.246$ kpc by using a modified (first-order) parameterized post-Newtonian formulation of General Relativity \citep{GRAVITY2020}. 

However, within this approach the authors do not compute the black hole angular momentum. Nevertheless, there are various techniques which can bound or estimate the rotation parameter of the black hole, for instance, by using flare emissions with a certain period, the value has been bounded by the following estimation $0.70 \pm 0.11 \, M \leq a \leq M$ \citep{trippe2007}, while an analysis that implements high-frequency quasi-period oscillation renders the following value $a \sim 0.996 M$ \citep{aschenbach1,aschenbach2,aschenbach3}. 
On the other hand, the estimated mass of the black hole hosted at the heart of M87, $M = 6.5\times 10^{9} M_{\bigodot}$, was obtained by using general relativistic magnetohydrodynamic models. 

In the near future, the BlackHoleCam project will join three different experiments: the Event Horizon Telescope \citep{EHT}, which will focus on the black hole event horizon looking for emissions from the relativistic plasma accreting onto SgrA*; GRAVITY \citep{gravity}, which will track the stars orbiting SgrA* with a near-infrared interferometer at the Very Large Telescope (VLT), and a set of radio telescopes (including ALMA) for detecting a radio pulsar in tight orbit about SgrA*. The main objective of the BlackHoleCam project is measuring with high accuracy the SgrA* parameters, assuming that a black hole is hosted in the center of our Galaxy. On the other hand, the Strong Gavity EU project \citep{SGEU2017} will analyze multi-wavelength spectral and fast timing observations of systems containing different kinds of black holes with the aim of estimating the mass and rotation parameters through measurements of X-ray radiation \citep{mass,spin}. Finally, the MICADO project at the Extremely Large Telescope \citep{trippe09} will analyze velocity profiles  and proper motions of surrounding the SgrA* stars and/or gas by spectroscopy with accuracies that improve the VLT/NACO ones about a factor of five, specifically the stellar proper motions of order of 10 $\mu as/yr$ (400 $m/s$) can be detected within a few years of observations with this facility.

Within General Relativity, the Kerr black hole hypothesis states that all isolated rotating astrophysical black holes are described by the Kerr spacetime and are completely defined by just two physical quantities: the mass $M$ and the spin parameter $a=J/M$, where $J$ is the black hole angular momentum in natural units, providing a remarkable prediction of the theory in the strong gravitational field regime. Within this paper we shall assume that the Kerr solution models real black holes.

Motivated by this fascinating research progress, in \citep{HAN} the authors proposed a novel general relativistic method for determining both the mass $M$ and the spin parameter $a$ of a Kerr black hole in terms of directly observed magnitudes with the smallest amount of assumptions and taking advantage of the conserved quantities of a stationary axisymmetric spacetime. Thus, these parameters were expressed in terms of the redshifts $z_{red}$ and blue-shifts $z_{blue}$ of emitted photons from geodesic particles (stars or gas) circularly orbiting around the Kerr black hole in the equatorial plane, as well as in terms of their orbital radii $r_e$. As a result, an eighth order polynomial equation for $M$ in terms of $z_{red}$, $z_{blue}$ and $r_e$ was obtained. This equation has no algebraic solution and one is forced to rely on a Bayesian fit to statistically determine the mass of the black hole. Once an estimation for the mass is obtained, the spin of the black hole can be easily computed through a simple closed expression.

In this work we push forward previous progress and find closed formulas for the Kerr black hole mass and spin parameters in terms of directly observed quantities (the red-, blue-  and central-shifts of emitted photons from the stars at three different points along their circular orbits and their orbital radii) by considering three different systems of geodesic particles circularly revolving around the black hole in the equatorial plane. 

The obtained analytical general relativistic expressions for the Kerr black hole parameters will be very useful when studying dynamical systems similar to the set of water masers revolving in an accretion disk around NGC 4258. For the case of stars orbiting out of the equatorial plane, like those revolving around SgrA*, our method needs a further development in order to render analytic formulas to estimate the black hole parameters. This is a work currently in progress.

In Sec. 2 and 3, we briefly review the results presented in \citep{HAN}, by considering the geodesics of massive test particles orbiting around a Kerr black hole and photons emitted by these bodies (by stars or gas), and by recalling the formulae for the kinematical red- and blue-shifts of emitted photons from stars/gas orbiting around a Kerr black hole.  In Sec 4, by studying a set of three stars revolving around the Kerr black hole along geodesic circular orbits in the equatorial plane, we obtain a linear algebraic equation for the mass parameter $M$ in terms of the red- and blue-shifts of photons emitted by these stars and their orbital radii; the corresponding analytical formulae for both the mass $M$ and the spin $a$ of the Kerr black hole are quoted. We similarly analyze a system of two geodesic particles with circular equatorial orbits around the black hole and obtain a cubic polynomial equation for $M$ with one real root, finding different general relativistic formulas for both $M$ and $a$ in terms of the stars' radii and the red- and blue-shifts of emitted photons by these bodies. We also solve the original problem presented in \citep{HAN} for a single particle circularly orbiting around the black hole, and obtain a new cubic polynomial equation for the mass parameter $M$ with an unique real root with the help of the central, red- and blue-shifts. Finally, in Sec. 5, we discuss our results and make some concluding remarks regarding the applicability of this general relativistic method to different astrophysical systems.


\section{ GEODESIC PARTICLES IN A KERR BH}

Here we shall review the geodesic motion of massive/massless particles in the Kerr background. The Kerr black hole family in Boyer-Linquist coordinates is given by the following metric:
\begin{equation}
\label{metric}
       ds^2 = g_{tt} dt^2 +  2 g_{t\varphi} dtd\varphi + g_{\varphi\varphi} d\varphi^2
       + g_{rr} dr^2 + g_{\theta\theta} d\theta^2,
\end{equation}
 where the components of the metric tensor $g_{\mu \nu}$ are:
\begin{eqnarray}
    g_{tt} & = & -\!\left(1-\frac{2Mr}{\Sigma}\right), \, \,  g_{t\varphi}=-\left(\frac{2Mar\sin^2 \theta}{\Sigma}\right),
    \, \,  g_{rr}=\frac{\Sigma}{\Delta},  \nonumber\\
        g_{\varphi\varphi} & =  & \left(r^2+a^2+\frac{2Ma^2r\sin^2 \theta}{\Sigma}\right)\sin^2\theta,
        \, \, 
    g_{\theta\theta}=\Sigma,  
\end{eqnarray}
where $\Delta = r^2 + a^2 - 2Mr,\,$ $\Sigma = r^2 + a^2 \cos^2 \theta$ and $M^2 \geq a^2 $.

The four-velocity of a geodesic test particle which moves in the Kerr gravitational field is given by $u^{\mu} = (u^{t}, u^{r}, u^{\theta}, u^{\varphi})$. If the four-velocity corresponds to a photon, then $u^{\mu}=k^{\mu}$, whereas if the four-velocity is associated to a massive test particle (a photons' emitter, like a star or gas), then $u^{\mu}=U^{\mu}$. These four-velocities are normalized, rendering the following condition:
$-\delta=u_{\mu}u^{\mu}$, where $\delta=0$ if the particles are photons and $\delta=1$ if they are massive particles.

Due to the existence of  time-like and rotation Killing vectors fields, we have two conserved quantities:
$E_{\delta} = - g_{\mu\nu} \xi^{\mu} u^{\nu}$,  and $L_{\delta} =  g_{\mu\nu} \psi^{\mu} u^{\nu}$. 
Here we define the energy $E\equiv E_{1}$ and the angular momentum $L \equiv L_{1}$ per unit mass of massive particles when $\delta=1$; and the energy $E_{\gamma}\equiv E_{0}$ and the angular momentum $L_{\gamma}\equiv L_{0}$ of emitted photons when $\delta=0$. 

From these relations one can obtain the effective potential for massive test particles \citep{Bardeen}:
\begin{eqnarray}
g_{rr} (U^{r})^2 \! + \! \underbrace{g_{\theta\theta} (U^{\theta})^2 \! + \! 1 \! - \!  \frac{E^{2} g_{\varphi \varphi}+
2E L\,g_{t\varphi}+L^2 g_{tt}}{(g_{t\varphi}^{2}-g_{tt}g_{\varphi\varphi})}}_{V_{eff}}=0.
    \label{Ur2}
\end{eqnarray}
The Kerr metric possesses a Killing tensor field that defines one more constant of motion
$C_{\delta} = K_{\mu\nu}u^{\mu}u^{\nu}$, 
which is related to the Carter constant $Q_{\delta}\,$ \citep{Carter} in the following way:
$C_{\delta} \equiv (L_{\delta}-aE_{\delta})^2 + Q_{\delta}.$ The Carter constant measures how much the path of the particles departs from the equatorial plane, vanishing when $\theta=\pi/2$. We further shall restrict ourselves to equatorial and circular trajectories ($U^{\theta}=0=U^r$) for simplicity, thus, the effective potential defined in (\ref{Ur2}) must obey the following conditions: $V_{eff}=0=V'_{eff}$, with the supplementary stability restriction $V''_{eff} > 0$ \citep{Bardeen}.

For equatorial circular orbits the following condition
\begin{equation}
\label{r_bound}
r > 2M\mp a+2\sqrt{M}\sqrt{M\mp a}
\end{equation}
must hold, whereas for stable orbits, the following one
\begin{eqnarray}
\label{r_ge}
&&r \geq M\left[ 3 + Z_2 \mp \sqrt{(3-Z_1)(3+Z_1+2Z_2)}\right], \\
&&Z_1=+\left(1-\frac{a^2}{M^2}\right)^{1/3}\left[ \left(1+\frac{a}{M}\right)^{1/3} + \left(1-\frac{a}{M}\right)^{1/3} \right], \nonumber   \\
&&Z_2=\sqrt{3a^2/M^2+Z_1^2},   \nonumber
\end{eqnarray}
where the $\pm$ signs correspond to co-rotating and counter-rotating photon sources, respectively.

Thus, the expressions for the non-trivial four-velocity components $U^{t}$ and $U^{\varphi}$ of stars become functions of the black hole parameters $M$ and $a$, and the orbital radius $r$ on the plane where the geodesic paths lie. Similar expressions hold for the photons' four-momentum $k^\mu$, subjected to the null normalization condition $k_\mu k^\mu=0$ \citep{HAN}.

\section{RED-/BLUE-SHIFTS OF PHOTONS EMITTED BY GEODESIC PARTICLES}

The frequency of a photon measured by an observer with four-velocity $U^{\mu}$ at a point $C$ reads
$\omega_C\equiv-k_\mu U^\mu_C \,$, this is a general relativistic invariant quantity, where the index $C$ indicates the emission $(e)$ or detection $(d)$ 
points of the measured photons' frequency. Thus, the general red-/blueshift in frequency that light signals 
emitted by massive particles experience in their path towards an observer, i.e. from the emission point 
($\omega_{e}$) till the detection point ($\omega_{d}$), is defined by $1+z = \frac{\omega_e}{\omega_d}$.
For circular equatorial star orbits this quantity becomes
\begin{equation}
\label{zcircorbits}
1+z =
\frac{\left.\left(E_\gamma U^t-L_\gamma
U^\varphi\right)\right|_e}{\left.\left(E_\gamma U^t - L_\gamma U^\varphi\right)\right|_d} =
\frac{U^t_e - b_e \,U^\varphi_e}{U^t_d - b_d \,U^\varphi_d}\,\,,
\end{equation}
where the impact parameter $b\equiv\frac{L_\gamma}{E_\gamma}$ was introduced.
This quantity is constant along the whole photons' path $b_e=b_d$ since the constants of motion $E_\gamma$ 
and $L_\gamma$ are preserved along the null geodesics followed by the photons from emission till detection. 

By considering the kinematic redshift of photons either side of the line of sight that links the Kerr black hole with the observer, one can subtract from Eq. (\ref{zcircorbits}) the expression for the redshift evaluated at the central value $z_c$ 
(see Fig.\,1), yielding
\begin{equation}
\label{zkin}
z_{\rm kin}\equiv z-z_c = \frac{U^t_e U^\varphi_d b_d - U^t_d U^\varphi_e b_e}
{U^t_d\left(U^t_d - b_d\,U^\varphi_d\right)},
\end{equation}
where $z_{c}=\left(U^t_e-U^t_d\right)/U^t_d$ is the redshift at the point $c$.

One further needs to consider the light bending generated by the gravitational field of the Kerr black hole, i.e. the mapping $b(r)$ 
between the impact parameter $b$ and the position of the photons' emitter particle $r$ in the equatorial plane. 
This quantity is maximized at the points where $k^r=0$. 
From the equality $k_\mu k^\mu=0$ one can obtain its expression for the Kerr metric restricted to the equatorial plane:
\begin{equation}
\label{bKerr}
b_{\pm} =\frac{-2aM \pm r \sqrt{r^2 + a^2 - 2Mr}}{r - 2M} \,,
\end{equation}
which yields two different values, $b_{\pm}$, that give rise to two different redshifts, $z$ and $z'$, of emitted photons corresponding to a receding/approaching source with respect to a given observer.

By considering that the detector is located far away from the photons' source $r_d\gg M\ge a$, the kinematic redshift of photons reads \citep{HAN}:
\begin{eqnarray}
\label{ZrKstatic}
z = \frac{\pm M^{\frac{1}{2}}\left(2aM + r_e\sqrt{r_e^2 - 2Mr_e + a^2\,}\right)}
{r_e^{\frac{3}{4}}\,\big(r_e-2M\big)\sqrt{r_e^{\frac{3}{2}}-3Mr_e^{\frac{1}{2}}\pm2aM^{\frac{1}{2}}\,}},\\
z' = \frac{\pm M^{\frac{1}{2}}\left(2aM - r_e\sqrt{r_e^2 - 2Mr_e + a^2\,}\right)}
{r_e^{\frac{3}{4}}\,\big(r_e-2M\big)\sqrt{r_e^{\frac{3}{2}}-3Mr_e^{\frac{1}{2}}\pm2aM^{\frac{1}{2}}\,}}.
\label{ZbKstatic}
\end{eqnarray}

 \begin{figure}[htb]
\centering
\includegraphics[scale=5]{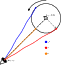} 
\caption{} \label{fig1}
\end{figure}
The quantities $z$ and $z'$ are the redshift and blueshift, respectively, when the test particles move in a co-rotating with the Kerr black hole way, and vice versa when the probe particles move counter-rotating.

Thus, the method consists in inverting the relations (\ref{ZrKstatic}) and (\ref{ZbKstatic}) in order to obtain expressions for the mass and rotation parameters of the Kerr black hole from the red and blue-shifts of photons emitted by stars in the special case of equatorial and circular orbits.

It is easy to obtain a closed expression for the spin parameter
\begin{eqnarray}
\label{a2}
a^2= \frac{\alpha\, r_e^3\,\left(r_e-2M\right)}
{4\,\beta\,M^2\, - \alpha\, r_e^2},
\end{eqnarray}
where $\alpha \equiv \left(z+z'\right)^2$ and $\beta \equiv \left(z-z'\right)^2$.  When substituting this relation into (\ref{ZrKstatic}) and (\ref{ZbKstatic}) one obtains an eighth order algebraic equation for the mass with no solution in terms of radicals and one must resort to a statistical fit in order to estimate this quantity from observational data. However, there is an alternative solution that we shall describe in the next Section.

\section{KERR MASS AND SPIN IN TERMS OF OBSERVED RED- AND BLUE-SHIFTS}

In this Section we shall express the Kerr black hole parameters $M$ and $a$ in terms of directly measured quantities, 
namely: the red-shifts $z$, $z'$ and $z_c$, and the orbital radius $r_e$ of the photons' source that revolve around the 
Kerr black hole in circular and equatorial orbits, accomplishing the aim stated above.

\subsection{A system with three orbiting particles}

We shall start by considering a system made-up of three probe particles that circularly rotate in the equatorial plane around the black hole with ordered orbital radii $r_1<r_2 <r_3$.

Then, for each of the test particles, labeled by the index $i$, the following equation must hold 
\begin{equation}
\label{gammai}
\gamma_i\equiv \frac{\alpha_i}{\beta_i}=\frac{4a^2 M^2}{r^2_i\left[ r^2_i+a^2-2Mr_i \right]}\,, \qquad i=1,2,3;
\end{equation}
where $\alpha_i=( z_i+z'_i)^2$ and $\beta_i= (z_i-z'_i)^2$. Thus, since each probe particle obeys its own Eq. (\ref{gammai}), 
in our approach we are considering that each test particle interacts only with the Kerr black hole, neglecting small corrections 
coming from the interactions among the particles themselves.  

Then, by dividing equation (\ref{gammai}) for the first(second) particle into that for the second(third) probe particle, and
solving for $a^2$ each of the resulting expressions, we obtain
\begin{equation}
\label{a12}
a^2=\frac{\gamma_2 r^3_2(r_2-2M)-\gamma_1 r^3_1(r_1-2M)}{\gamma_1 r^2_1-\gamma_2 r^2_2}\,,
\end{equation}
\begin{equation}
\label{a23}
a^2=\frac{\gamma_3 r^3_3(r_3-2M)-\gamma_2 r^3_2(r_2-2M)}{\gamma_2 r^2_2-\gamma_3 r^2_3}\,.
\end{equation}
By equating these relations we 
obtain a linear equation for $M$, which allows us to get a general relativistic closed formula for the mass parameter 
in terms of the red- and blue-shifts and the orbital radii of all the particles involved in the configuration:
\begin{eqnarray}
\label{Mas}
 M  =\frac{1}{2} \frac{ \sum_{i < j}(-1)^{i+j+1} \gamma_i \gamma_j r^2_i r^2_j (r^2_j-r^2_i) }{ \sum_{i < j}(-1)^{i+j+1} \gamma_i \gamma_j r^2_i r^2_j (r_j-r_i)}\, .
\end{eqnarray}
Once the calculation for the Kerr mass is made with the aid of (\ref{Mas}), the resulting value should be inserted in either Eq. (\ref{a12}) or Eq. (\ref{a23}) in order to get a general relativistic expression for the value of the spin parameter of the Kerr black hole.
 
Thus, with this method we have managed to write general relativistic closed expressions for the mass $M$ and spin $a$ parameters of a Kerr black hole in terms of twelve observational measurements of three orbiting stars: three different redshifts for each particle and their three orbital radii. 

Remarkably, the reduction from eighth to first order in the algebraic equation for $M$ was accomplished by considering a system with more geodesic particles orbiting around the Kerr black hole, conveniently involving more observational data.

\subsection{A system with two orbiting particles}

For this system it will be useful to define $r_1 \equiv r$ and $r_{2} \equiv  \lambda r$ such that the constant $ \lambda \equiv \frac{r_2}{r_1} >1$ since $r_2>r_1$.

Then, the expressions given by Eq. (\ref{a2}) for the first and second orbiting test particles respectively read
\begin{equation}
\label{a1}
a^2= \frac{\alpha_{1} r^3(r-2M)}{4 \beta_{1} M^{2} - \alpha_{1} r^2}
=\frac{\alpha_{2} (\lambda r)^3(\lambda r-2M)}{4 \beta_{2} M^{2} - \alpha_{2} (\lambda r)^2}\,.
\end{equation}
By combining these equations we obtain a third order algebraic equation for the ratio $M/r$:
\begin{eqnarray}
\label{M1}
8(\alpha_1 \beta_2 & - &  \lambda^3 \alpha_2 \beta_1  )\frac{M^3}{r^3}-4(\alpha_1 \beta_2 - \lambda^4 \alpha_2 \beta_1 )\frac{M^2}{r^2}\nonumber \\ 
 &+& 2 \lambda^2 (\lambda -1) \alpha_1 \alpha_2 \frac{M}{r}- \lambda^2 (\lambda^2 -1) \alpha_1 \alpha_2 =0\,.
\end{eqnarray}
It can be shown that the discriminant $\Delta_3$ of Eq. (\ref{M1}) is negative since the cubic coefficient is greater than the quadratic one. 
Therefore, there is only a real root for the ratio $M/r$ and the solution for the mass can be expressed in the following way:
\begin{equation}
\label{M11}
M=\frac{r}{6\left(\alpha_1\beta_2-\lambda^3 \alpha_2\beta_1 \right )}\Big[ \alpha_1\beta_2-\lambda^4 \alpha_2\beta_1+\chi+\frac{\Psi}{\chi}\Big],
\end{equation}
where we have introduced the following relations  
\begin{eqnarray}
\chi&=&\left[4 \left( \Phi+\sqrt{\Phi^2-4\Psi^3} \right)\right]^{\frac{1}{3}}, \, \, Phi=\Upsilon +\sqrt{4 \Psi^3+\Upsilon^2}\,,  \nonumber  \\
\Psi&=&\alpha^2_1 \beta_2 \left[3 \lambda^2(\lambda-1) \alpha_2-\beta_2 \right]+2\lambda^4 \alpha_1 \alpha_2 \beta_1 \beta_2\nonumber \\ 
&-&\lambda^5 \alpha^2_2 \beta_1  \left[ 3(\lambda-1)\alpha_1+ \lambda^3 \beta_1 \right],  \nonumber   \\
\Upsilon&=&\beta^2_2 \alpha^3_1 \left[ 2\beta_2+9(3 \lambda+2)(\lambda-1)\lambda^2\alpha_2 \right]-3\lambda^2 \alpha_1 \alpha_2 \beta_1\beta_2  \nonumber\\
 &\times &\left[ 2 \lambda^2 (\alpha_1 \beta_2+\lambda^2 \alpha_2 \beta_1) +3(\lambda^2-1)(\lambda^3+6)\alpha_1 \alpha_2 \right]\nonumber \\
 &+&\lambda^8 \alpha^3_2 \beta_1 \left[ 9(\lambda-1)(3\lambda+2)-\lambda^4 \beta_1 \right]\,,   \nonumber
\end{eqnarray}
providing an analytic general relativistic formula for the mass of the Kerr black hole in terms of few observational data.

With this expression at hand, it is straightforward to obtain a closed general relativistic formula for the spin parameter of the Kerr black hole:
\begin{eqnarray}
\label{a2o}
a^2\!=\!
\frac{ 3 \alpha_1 \epsilon \chi^{1/3} \! \left[ \chi^{2/3}-\left(\epsilon+\lambda^4 \alpha_2 \beta_1 \! \right) \chi^{1/3}-2 \Psi \right] r^2 }
{9 \alpha_1 \epsilon^2 \chi^{2/3} \! - \! \beta_1 \! \left[\chi^{2/3} \! + \! 2(\alpha_1 \beta_2 \! - \!\lambda^4 \alpha_2 \beta_1) \chi^{1/3} \!\! - \! 2 \Psi \right]^2},  
\end{eqnarray}
where $\epsilon = \alpha_1 \beta_2-\lambda^3 \alpha_2 \beta_1$. Here we would like to stress that it is not relevant at all if a given star/gas is co-rotating or counter-rotating with the Kerr black hole due to the quadratic dependence on the spin parameter $a$ in Eqs. (\ref{a1}). Therefore, this method allows us to obtain just the magnitude of the spin parameter. Thus, we managed to write down general relativistic closed formulae for the mass $M$ and spin $a$ parameters of a Kerr black hole in terms of eight observational measurements of two orbiting stars: three redshifts per particle and their orbital radii. It is quite remarkable that we reduced an eighth order algebraic equation for $M$ with four required observational data to a cubic one with just eight needed observational measurements.


\subsection{A single orbiting particle}

We finally approach and solve the original problem that was posed in \citep{HAN}. In order to find an expression for the Kerr mass in the case of a single orbiting particle, we recall the definition of the central redshift:
\begin{eqnarray}
\label{zc1}
1+z_{c}= \frac{r^{\frac{3}{2}}\pm a M^{\frac{1}{2}}}{r^{\frac{3}{4}}\sqrt{r^{\frac{3}{2}}-3Mr^{\frac{1}{2}}\pm 2 a M^{\frac{1}{2}}}}, 
\end{eqnarray}
define the square of Eq. (\ref{zc1}) as $\kappa\equiv(1+z_c)^2$ and 
the product of expressions (\ref{ZrKstatic}) and (\ref{ZbKstatic}) in the following way
\begin{equation}
\label{sigma}
\sigma \equiv z z' = -M\frac{r^3-(r+2M)a^2}{(r-2M)(r^{3}-3M r^{2} \pm 2 a r^{\frac{3}{2}} M^{\frac{1}{2}})}.
\end{equation}
Then, by clearing the term  $\pm 2 a r^{\frac{3}{2}} M^{\frac{1}{2}}$ from $\kappa$ and $\sigma$ we obtain
\begin{equation}
\label{woa2}
[(\kappa-1)(r+2M)-\sigma (r-2M)]   a^2  = 3 \sigma (r-2M) r^2 +(\kappa -1)r^3;  
\end{equation}
further substitution of the expression (\ref{a2}) for $a^2$ into this relation yields a third order polynomial equation for M in terms of purely observational data:
\begin{eqnarray}
\label{poly3}
12 \beta \sigma M^3 & - & 2\left[ (\sigma+\kappa-1)(\alpha+\beta) + 2\beta\sigma\right] M^2 r - \nonumber \\ 
\alpha \sigma M r^2 & + & \alpha(\kappa-1-\sigma)r^3=0. 
\end{eqnarray}
By writing this relation as
\begin{eqnarray}
\label{poly3eje}
12p_3 \left(\frac{M}{r} \right)^3 -2p_2\left(\frac{M}{r}\right)^2-p_1\frac{M}{r}+p_0=0,   \nonumber
\end{eqnarray}
where $p_3$, $p_2$, $p_1$ and $p_0$ are the cubic, quadratic, linear and zero degree coefficients of Eq. (\ref{poly3}), 
respectively. It is easy to find, with the help of the inequalities (\ref{r_bound}) and (\ref{r_ge}), that:
\begin{eqnarray}
\label{assu}
 |p_2|>|p_3|>p_0>|p_1|.   \nonumber
 \end{eqnarray}
With the aid of these relations it can be shown that the corresponding discriminant of the Eq. (\ref{poly3}) is negative, $\Delta_3<0$, leading to an unique real root given by:
\begin{equation}
\label{root}
M=\frac{r}{18\beta \sigma}\left[(\sigma+\kappa-1)(\alpha+\beta)+2\beta\sigma+\Lambda+
\frac{\Gamma}{\Lambda}\right], 
\end{equation}
where we have introduced the following expressions:
\begin{eqnarray}
\Lambda &=& \left(\frac{\Omega+\sqrt{\Omega^2-4\Gamma^3}}{2} \right)^\frac{1}{3}, \nonumber \\
\Gamma &=& \left[(\sigma+\kappa-1)(\alpha+\beta)+2\beta\sigma\!\right]^2-9 \alpha\beta\sigma^2,  \nonumber \\
\Omega &=& 2\left[(\sigma+\kappa-1)(\alpha+\beta)+2\beta \sigma \right]^3-486\alpha\beta^2 
\sigma^2(\kappa-1-\sigma)
\nonumber \\ 
&+&27 \alpha \beta \sigma^2 \left[(\sigma+\kappa-1)(\alpha+\beta)+2\beta \sigma \right].  \nonumber 
\end{eqnarray}
With the help of Eq. (\ref{root}) we obtain a general relativistic closed formula for the spin parameter $a$ in the following way:
\begin{eqnarray}
\label{aspin}
a^2=18\alpha \sigma \Lambda\frac{18\alpha \beta \sigma \Lambda+\Xi}{\Xi^2-386\alpha \beta \sigma^2 \Lambda^2},
\end{eqnarray}
where $
\Xi=\Lambda \left[ (\sigma+\kappa-1)(\alpha+\beta)+2\beta \sigma+\Lambda \right] + \Gamma.$

It is worth noticing that in this single orbiting star case, we have reduced the eighth order polynomial equation for $M$ presented in \citep{HAN} to a cubic one with the same number of observational data. The only difference with the previous method (in contrast with the aforementioned three and two stars' cases) is that the information of the central shift $z_c$ enters explicitly in the final formulas for the single geodesic particle case.


\section{CONCLUDING REMARKS}

In this work we managed to derive general relativistic closed expressions for the mass $M$ and spin $a$ parameters of a Kerr black hole 
in terms of twelve (eight) observational measurements of three (two) orbiting stars: three different redshifts for each particle and 
their three (two) orbital radii. We obtained as well closed formulas for these parameters in the single particle case by making explicit use  of the central shift $z_c$.  

Here we would like to emphasize that this approach analyzes the black hole rotation curves on the basis of directly measured general relativistic 
invariant quantities: the gravitational red- and blue-shifts, in contrast to the radial velocities, which are coordinate dependent observables.
Moreover, when suitable, astronomers can make use of all the proposed configurations with a different number of orbiting stars in order to reduce 
the statistical errors involved in the configuration when just one star is orbiting around the black hole.

This method can also be used as a null test of the Kerr black hole hypothesis since the expressions for $z$ and $z'$ are bounded: 
if the red-/blue-shifts observational data of a given set of stars orbiting around a black hole do not fall within the range predicted by 
General Relativity, then the black hole will not be of Kerr type, opening the possibility for black hole solutions of modified theories 
of gravity to describe the dynamics of the above mentioned revolving stars (see \citep{SheoranHAN} and references therein).

Finally, a simplified version of this method has been used to estimate the black hole mass of the dynamical system NGC 4258 in \citep{NHALCLC}, while a suitable modification of it can be applied to the S0 set of stars orbiting the black hole SgrA*, to other galactic center with supermassive black holes, like the core of M31, where two red giants (known as P1 and P2) whose kinematics are consistent with circular stellar disks, revolve around a supermassive black hole known as P3 \citep{M31} with mass  $M_\bullet \approx 1.4 \times 10^{8} M_{\bigodot}$, and to Active Galactic Nuclei like Centaurus A (NGC5128) where both molecular gas \citep{Neumayer:2007ic} and red stars \citep{CenA} orbit around a supermassive black hole of mass  $M_\bullet \approx 5 \times 10^{7} M_{\bigodot}$.

{\bf Acknowledgments} A.H.-A. thanks the Raman Research Institute for hospitality and is grateful to S. Sethi, Sridhar S., B. Paul, B. Mukhopadhyay, J. Samuel, J.A. Olvera-Santamar\'ia and A. Gonz\'alez-Ju\'arez for fruitful and insightful discussions on this topic. A.H.-A. and U.N. thank SNI and acknowledge support from VIEP-BUAP and CIC-UMSNH. All authors acknowledge financial support from CONACYT: A.H.-A.'s grant No. A1-S-38041, R. L.-C.'s grant No. 15706, and U.N.'s grant No. 2558591. A.H.-A. andU.N. thank the thematic network project 280908 {\it `Agujeros Negros y Ondas Gravitatorias'}.

\bibliography{Aguilar}%

\end{document}